\documentclass[prl,twocolumn,showpacs,amsmath,amssymb,superscriptaddress]{revtex4-1}

\usepackage{graphicx}% Include figure files
\usepackage{color}
\usepackage{subfigure}
\usepackage{dcolumn}% Align table columns on decimal point
\usepackage{bm}% bold math
\usepackage{epsf}
\usepackage{amsmath,amstext,amssymb}
\usepackage{enumitem}
\usepackage{hyperref}
\usepackage[ansinew]{inputenc}
\usepackage{braket}
\usepackage{bbold}
\usepackage{siunitx}
\usepackage{lipsum}
\usepackage{array}
\usepackage{float}
\PassOptionsToPackage{numbers,sort&compress}{natbib}
\usepackage{pdfpages}%for attaching the supplemental
\makeatletter
\AtBeginDocument{\let\LS@rot\@undefined}
\makeatother

\usepackage{hyperref}
\hypersetup{
colorlinks=true,
linkcolor=blue,
citecolor=blue,}
\usepackage{textcomp}
\usepackage{tikz}
\usetikzlibrary{quantikz}
\makeatletter
\g@addto@macro\normalsize{%
	\setlength\abovedisplayskip{4pt}
	\setlength\belowdisplayskip{4pt}
	\setlength\abovedisplayshortskip{4pt}
	\setlength\belowdisplayshortskip{4pt}
}
\makeatother

\newcommand{\be}{\begin{equation}}
\newcommand{\ee}{\end{equation}}
\newcommand{\bea}{\begin{eqnarray}}
\newcommand{\eea}{\end{eqnarray}}

\newcommand{\St}{{\cal S}}
\newcommand{\LL}{{\cal L}}
\newcommand{\expec}[1]{\left\langle #1 \right\rangle}
\usepackage{mhchem}

\begin{document}

\title{Fault-tolerant parity readout on a shuttling-based trapped-ion quantum computer}

\author{J.~Hilder}
\affiliation{Institut f\"ur Physik, Universit\"at Mainz, Staudingerweg 7, 55128
	Mainz, Germany}
\author{D.~Pijn}
\affiliation{Institut f\"ur Physik, Universit\"at Mainz, Staudingerweg 7, 55128
	Mainz, Germany}
\author{O.~Onishchenko}
\affiliation{Institut f\"ur Physik, Universit\"at Mainz, Staudingerweg 7, 55128
	Mainz, Germany}
\author{A.~Stahl}
\affiliation{Institut f\"ur Physik, Universit\"at Mainz, Staudingerweg 7, 55128
	Mainz, Germany}
\author{M.~Orth}
\affiliation{Institut f\"ur Physik, Universit\"at Mainz, Staudingerweg 7, 55128
	Mainz, Germany}
\author{B.~Lekitsch}
\affiliation{Institut f\"ur Physik, Universit\"at Mainz, Staudingerweg 7, 55128
	Mainz, Germany}
\author{A.~Rodriguez-Blanco}
\affiliation{Departamento de F\'isica Te\'orica, Universidad Complutense, Madrid 28040, Spain}
\author{M.~M\"uller}
\affiliation{Institute for Quantum Information, RWTH Aachen University, D-52056 Aachen, Germany}
\affiliation{Peter Gr\"unberg Institute, Theoretical Nanoelectronics, Forschungszentrum J\"ulich, D-52425 J\"ulich, German}
\author{F.~Schmidt-Kaler}
\affiliation{Institut f\"ur Physik, Universit\"at Mainz, Staudingerweg 7, 55128
	Mainz, Germany}
\author{U.~G.~Poschinger}\email{poschin@uni-mainz.de}
\affiliation{Institut f\"ur Physik, Universit\"at Mainz, Staudingerweg 7, 55128
	Mainz, Germany}

\begin{abstract}
Quantum error correction requires the detection of errors by reliable measurements of suitable multi-qubit correlation operators. Here, we experimentally demonstrate a fault-tolerant weight-4 parity check measurement scheme. An additional 'flag' qubit serves to detect errors occurring throughout the parity measurement, which would otherwise proliferate into uncorrectable weight-2 errors on the qubit register. We achieve a flag-conditioned parity measurement single-shot fidelity of  93.2(2)\%. Deliberately injecting bit and phase-flip errors, we show that the fault-tolerant protocol is capable of reliably intercepting such faults. For holistic benchmarking of the parity measurement scheme, we use entanglement witnessing to show that the implemented circuit generates genuine six-qubit multi-partite entanglement. The fault-tolerant parity measurement scheme is an essential building block in a broad class of stabilizer quantum error correction protocols, including topological color codes. Our hardware platform is based on atomic ions stored in a segmented microchip ion trap. The qubit register is dynamically reconfigured via shuttling operations, enabling effective full connectivity without operational cross-talk, which provides key capabilities for scalable fault-tolerant quantum computing.
\end{abstract}

\maketitle

\section{Introduction}

Quantum computers promise to outperform classical processors for particular tasks \cite{nielsen00, Montanaro2016, Arute2019, PRXQuantum.2.017001}. 
Solving problems beyond the reach of classical computers with a universal quantum computer requires the implementation of quantum error correction (QEC) protocols \cite{Terhal_2015} to mitigate faulty operational building blocks. In QEC codes, logical qubits are encoded into entangled states of several physical qubits. Error syndrome readout permits detection of errors through quantum non-demolition (QND) parity check measurements (PCM) on the logical qubits \cite{PhysRevLett.99.120502, Lupascu2007, Barreiro2011, Corcoles2015}. A QND PCM requires performing a sequence of entangling gates between a set of data qubits and an ancilla qubit, to which the parity information is mapped \cite{Devitt_2013}. Projective measurements on the ancillae discretize eventual errors and thus allow for their detection and subsequent correction. However, PCM circuits consist of faulty gate operations and may therefore corrupt the qubit register. Therefore, \textit{fault-tolerant} (FT) QEC schemes are needed to prevent uncontrolled proliferation of errors through the quantum registers \cite{shor1996faulttolerant}.
\begin{figure}[h!tp]\begin{center}
\includegraphics[width=\columnwidth,trim={2cm 0cm 3cm 0cm},clip]{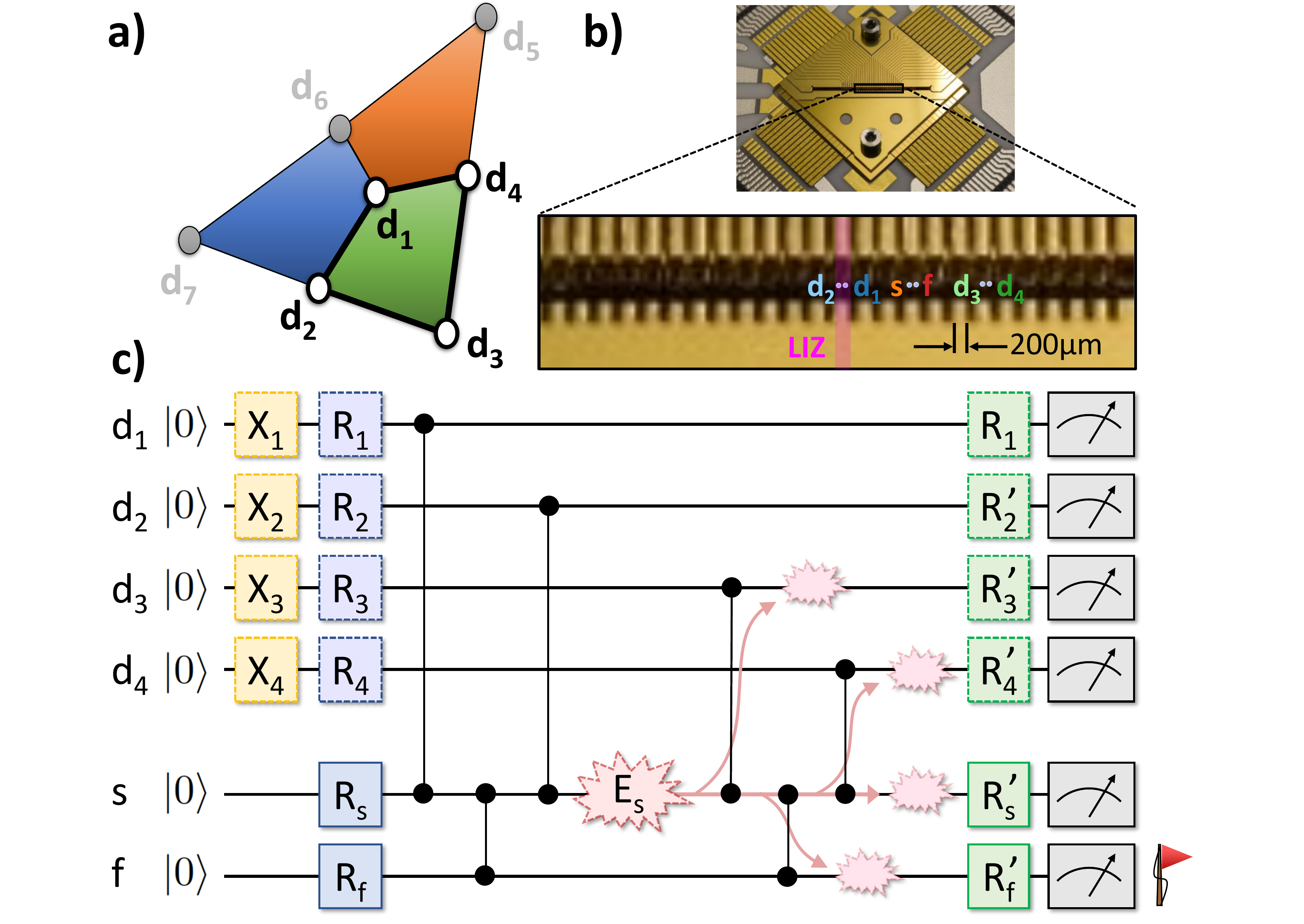}
\caption{\textbf{a)} Sketch of the topological $[[7,1,3]]$ color code, highlighting a plaquette comprised of data qubits $d_1-d_4$. \textbf{b)} Segmented microchip ion trap used in this work, with the enlarged picture showing how the six ion qubits are distributed at the beginning of the gate sequence. \textbf{c)} Quantum circuit for a fault-tolerant parity check measurement on four data qubits. Entangling gates map the parity of the data qubits to the syndrome qubit $s$, while two additional gates serve for detecting potentially uncorrectable faults using the flag qubit $f$. An error $E_s$ (red) propagates through subsequent gates and results in a weight-2 error on the data qubit register is detected by the flag qubit. Initialization (yellow, blue) and analysis (green) rotations are carried out on all qubits at the beginning of the sequence and immediately before projective readout (grey). 
}
 \label{fig:circuit}
\end{center}\end{figure}
Previously conceived FT PCM schemes demand adding as many ancilla qubits as the parity check generator with maximum weight \cite{PhysRevLett.77.3260,PhysRevLett.98.020501}. More recent FT PCM schemes, based on so-called flag-qubits, substantially reduce the overhead in terms of qubits and gate operations \cite{PhysRevLett.121.050502, Chamberland2018flagfaulttolerant, Chamberland2019faulttolerantmagic,PhysRevA.100.062307, Reichardt_2020, PhysRevA.101.012342, Chamberland_2020,PhysRevA.101.012342,PRXQuantum.1.010302}. In particular, for distance-three codes implemented in fully connected quantum registers, a total of only two ancilla qubits is sufficient to maintain the one-fault detection and correction condition \cite{Chamberland2018flagfaulttolerant}, i.e.~to guarantee the correctability of one arbitrary error occurring on any of the qubits or operations involved in the logical qubit.\\

To date, several QEC protocols and components have been demonstrated, using trapped ions \cite{NIGG2014,PhysRevX.6.031030,SchindlerScience1059,kielpinski2001recent,Chiaverini2004,Negnevitsky2018,Stricker2020}, superconducting circuits \cite{Kelly2015, ofek2016demonstrating, Andersen2020,chen2021exponential}, nuclear magnetic resonance \cite{s-zhang-prl-109-100503, s-knill-prl-86-5811}, or nitrogen-vacancy centers \cite{Waldherr_2014,Unden2016}. Increasing gate fidelities for different platforms \cite{Srinivas2021, CraigClark2021,kjaergaard2020programming,Browaeys2020} render QEC circuit noise thresholds \cite{raussendorf-prl-98-190504} to be within reach of experimental capabilities. So far, with regard to FT QEC elements, FT state preparation and detection on %primitive parts 
primitives of topological surface codes have been realized with superconducting circuits \cite{Corcoles2015, PhysRevLett.119.180501,Kelly2015} or trapped ions \cite{Linkee1701074}. Recently, FT preparation and FT operations of an encoded qubit on a distance-3 Bacon-Shor code was demonstrated \cite{egan2021faulttolerant}, where the FT syndrome extraction was realised using four ancilla qubits in addition to the nine data qubits. \\

\subsection{Fault tolerant parity check measurement}
In this work, we employ a trapped ion quantum processor to demonstrate a flag-based FT weight-4 PCM scheme, which reduces the overhead for FT syndrome readout to two extra \textit{syndrome} and \textit{flag} qubits. The flag qubit detects \textit{hook errors}, i.e. faults occurring on the syndrome qubit that proliferate onto two errors on the data qubit register. They would remain undetectable in a non-FT PCM scheme and eventually result in a logical error. In general, a weight-4 FT PCM circuit represents a key building block of the smallest distance-3 topological color code, which is equivalent to the $[[n=7,k=1,d=3]]$ Steane code \cite{PhysRevLett.77.793, PhysRevLett.97.180501}, as well as of FT circuit constructions for larger 2D topological QEC codes \cite{Chamberland2018flagfaulttolerant, PhysRevA.101.012342, PhysRevA.101.032333}. This stabilizer code \cite{stabilisers} encodes $k=1$ logical qubit into $n=7$ physical qubits with a code distance $d=3$ and can therefore correct up to $t=(d-1)/2=1$ arbitrary error on any of the physical qubits, provided that QEC cycles are realized via fault-tolerant circuit constructions, based e.g.~on the flag-qubit based FT PCM measurement demonstrated in this work. The physical qubits of the code can be arranged in a 2D triangular lattice structure formed by three interconnected 4-qubit plaquettes, as displayed in Fig.~\ref{fig:circuit}a. The set of parity check or stabilizer generators $\{g_i\}$ of the code generate the stabilizer group $\St$ 
and are 4-qubit Pauli operators defined on vertices $v(p)$ of each its plaquettes $p$: 
\begin{equation}
g_x^{(p)}=\bigotimes_{i\in v(p)}X_i,\qquad g_z^{(p)}=\bigotimes_{i\in v(p)}Z_i,
\end{equation}
with the Pauli matrices $X_i,Y_i,Z_i,\mathbb{1}_i$ pertaining to qubit $i$.
The \textit{code space} $\LL$ hosting the logical qubit is fixed as the common two-dimensional eigenspace of eigenvalue $+1$ of all generators $g_i$ (and combinations thereof), 
\begin{equation}
   \LL:= \{\ket{\psi}_{\LL}:g_i\ket{\psi}_{\LL}=+\ket{\psi}_{\LL}\quad \forall g_i %\in\St
   \}.
\end{equation}

Here, we focus on the experimental verification of a flag-based FT weight-4 parity check, $g_z=Z_1Z_2Z_3Z_4$, according to the circuit shown in Fig.~\ref{fig:circuit}c. The $g_x$ parity check is equivalent, as it merely requires mapping the data qubits by local rotations to the $X$ basis before syndrome readout. Four entangling gates of type $Z_i\otimes Z_j$ lead to a $\pi$ phase shift on the syndrome for odd parity of the four data qubits, which is detected upon readout. Two additional entangling gates between the syndrome and flag qubits serve for catching error events throughout the PCM, which would otherwise result in weight-2 errors on the data qubit register (see Fig. \ref{fig:circuit} c).

QEC is intimately linked to multipartite entanglement \cite{Preskilnotes, PhysRevA.54.3824, entanglementassistQECC,Almheiri2015,PRXQuantum.2.020304}. There are several works that reveal explicit connections between QEC and the production of maximally entangled states and equivalently, between entanglement fidelities of the encoded states and the weight distribution of a code \cite{PhysRevA.69.052330,Raissi_2018,PhysRevLett.78.1600,681316}. The inherent relation of non-classical correlations as a prerequisite for QEC renders the generation and verification of genuinely multipartite entangled (GME) states to be a suitable benchmarking protocol for FT QEC building blocks. Here, we verify GME between the data and ancilla qubits in order to demonstrate the correct functioning of our FT PCM, and to benchmark the capabilities of our trapped-ion processor in the context of FT QEC.

\section{Shuttling-Based Trapped-Ion Platform}

Quantum computer platforms based on trapped atomic ions arranged as static linear registers and laser addressing have seen substantial progress \cite{Blatt14Qubit,Debnath2016,Linke3305}. On such platforms, QEC building blocks have been demonstrated, such as repeated syndrome extraction and correction \cite{Schindler1059,Negnevitsky2018}, encoding, readout and gate operations for the $[[7,1,3]]$ code \cite{Nigg302}, and entanglement of encoded logical qubits \cite{Erhard2021}. However, QEC protocols impose stringent demands on the scalability of the underlying hardware platform. The shuttling-based ``Quantum-CCD''  approach offers a route to increased scalability \cite{KIELPINSKI2002,Lekitsch2017,Kaushal2020,Pino2021}. Here, the qubit ions are kept in the form of small subsets within a microstructured trap array, and the register is dynamically reconfigured via shuttling operations. This way, the excellent degree of control can be retained for increasing register sizes. In this work, we implement a shuttling-based FT PCM protocol. Between subsequent gate operations on two qubits, the register is reconfigured via shuttling operations. A special feature of our protocol is that we establish the required effective all-to-all connectivity by reordering the register via physical rotation of two commonly confined ions. This operation is equivalent to a unit-fidelity SWAP logic gate \cite{SWAPGATE} and contrasts with faulty radiation-driven SWAP gates. This, together with the inherently low cross-talk of the shuttling-based architecture, allows to maintain the one-fault QEC condition.\\

\begin{figure*}[h!tp]\begin{center}
\includegraphics[width=\textwidth,trim={0cm 0.3cm 0cm 0cm},clip]{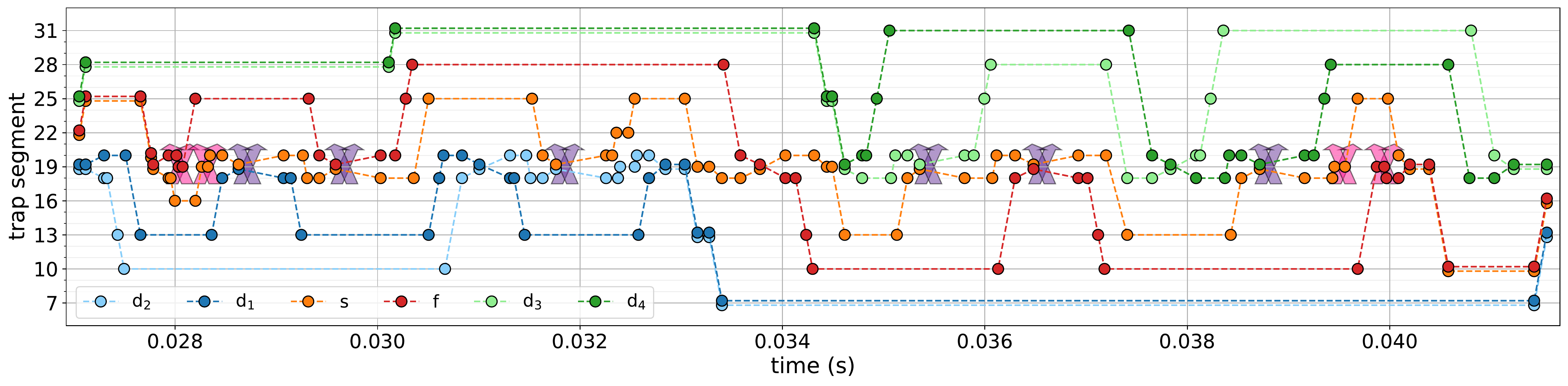}
\caption{Shuttling schedule of the fault-tolerant parity readout measurement sequence, indicating how the ion qubits are moved between different storage sites of the segmented ion trap. The fixed laser interaction zone is located at segment 19. The arrows indicate laser-driven gate interactions. A distance of at minimum two empty segments between sets of ion qubits is maintained throughout the sequence. The maximum spatial extent of the register is 24 segments (4.8~mm).}
\label{fig:6IonFTRShuttling}
\end{center}
\end{figure*}

We employ a micro-structured, segmented radio frequency ion trap \cite{Kaushal2020}, consisting of 32 uniform segment pairs which are linearly arranged along a \textit{trap axis}. Each segment pair can generate a confining potential well. All qubit operations - initialization, gates and readout - are carried out using laser beams, which are directed to segment 19 - henceforth referred to as \textit{laser interaction zone (LIZ)}. Shuttling operations are carried out by supplying suitable voltage waveforms to the trap electrodes. Potential wells containing one or two qubits can be moved along the trap axis \cite{WALTHER2012,BOWLER2012}, two commonly confined ions can be separated into potential wells \cite{RUSTER2014,KAUFMANN2014}, two separately confined ions can be merged into one well, and two commonly confined ions can be rotated such that their ordering along the trap axis is reversed \cite{SWAPGATE}. The \textit{separate / merge} and \textit{swap} shuttling operations are limited to the LIZ.\\
The qubits are encoded in the spin of the valence electron of atomic $^{40}$Ca$^+$ ions \cite{POSCHINGER2009,RusterLongLived2016}, with the assignment $\ket{0}\equiv \ket{S_{1/2},m_J=+1/2}, \ket{1}\equiv \ket{S_{1/2},m_J=-1/2}$.  Gate operations are driven by a pair of beams detuned by about $2\pi \times 1.0$~THz from the $S_{1/2}\leftrightarrow P_{1/2}$ electric dipole transition. Local qubit rotations are realized via stimulated Raman transitions, allowing for arbitrary local rotations on qubit $i$ of the form
\begin{equation}
    R_i(\theta,\phi)=\exp\left[-i\frac{\theta}{2}\left(\cos\phi\;X_i+\sin\phi\;Y_i\right)\right],
\end{equation}
Local rotations can also be carried out simultaneously on two qubit ions commonly confined in the LIZ, in which case the Pauli operators are to be replaced by the respective tensor sum operators. Entangling gates between any two qubits $i$ and $j$ are realized via spin-dependent optical dipole forces \cite{LEIBFRIED2003A}, effecting a phase shift $\Phi$ between even and odd parity (with respect to the $Z$ basis) states, represented by the unitary
\begin{eqnarray}
    ZZ_{ij}(\Phi)&=&e^{\frac{i}{2}\Phi Z_i \otimes Z_j }.
    \label{eq:entanglinggatesPhi}
\end{eqnarray}
We employ a maximally entangling gate with total phase $\Phi = \pi/2$, accumulated from two separate gate pulses, interspersed by a rephasing $\pi$ pulse. This leads to the total gate unitary
\begin{eqnarray}
    G_{ij}&=&ZZ_{ij}(\pi/4)R(\pi,-\pi/2)ZZ_{ij}(\pi/4).
    \label{eq:entanglinggates}
\end{eqnarray}
The rephasing pulses serve to maintain coherence \cite{Biercuk2009}, especially on the syndrome and flag qubits, which undergo multiple entangling gates. Upon gate operations, the potential well in the LIZ features single-ion secular frequencies of $2\pi\times\{1.49,3.88,4.64\}$~MHz, with the lowest frequency pertaining to the trap axis. Entangling gates are carried out using the transverse in-phase collective vibrational mode at $2\pi \times 4.64$~MHz as the gate-mediating mode. The laser beam geometry is chosen such that the gate operations are insensitive to the collective modes oscillating along the trap axis, which accumulate excitation from shuttling operations \cite{PhysRevLett.119.150503}.   \\

The shuttling schedule realizing the FT PCM is constructed from the primitive operations described above, such that the total count of shuttling operations and the maximum spatial extent of the register is minimized, while additional constraints such as the minimum number of empty trap segments between two qubit sets are always fulfilled. Initially, the qubits are stored pairwise in order $\{d_2,d_1\},\{s,f\}$ and $\{d_3,d_4\}$. The ion pairs are sequentially moved to the LIZ, where all four transverse modes are cooled close to the ground state via resolved-sideband cooling \cite{POSCHINGER2009}, and the qubits are initialized to $\ket{0}$ via optical pumping. Then, the data qubit sets $\{d_2,d_1\},\{d_3,d_4\}$ are moved to the LIZ, where they are separated. Each data qubit is again moved into the LIZ, where optional $\pi$-flips allow for preparation of any desired logical basis state. A similar procedure is then  carried out for the syndrome and flag qubits, which are  prepared in superposition states via $\pi/2$ rotations. Then, the parity mapping sequence is carried out: the qubits undergo pairwise entangling gates according to Eq. (\ref{eq:entanglinggates}) in the sequence $d_1s,sf,d_2s,d_3s,sf,d_4s$. Before and after each gate, an optional $\pi/2$ rotation can be carried out on the participating data qubit in order to change the basis. Between two consecutive gates, a sequence of movement, separate / merge and position-swap operations is carried out, bringing the qubit pair on which the following gate operation is to be carried out to the LIZ, see Fig. \ref{fig:6IonFTRShuttling}.  Upon completion of the gate sequence, the syndrome and qubits are separately moved to the LIZ and each undergo an analysis $\pi/2$ rotation. Qubit phases accumulated from positioning in the inhomogeneous magnetic field have been calibrated via previous Ramsey-type measurements \cite{WALTHER2012,PhysRevX.7.031050} and are corrected for. Upon completion of the gate sequence, the qubits are kept pairwise in order $\{d_2,d_1\},\{s,f\}$ and $\{d_3,d_4\}$. These pairs are sequentially moved to the LIZ in reverse order, where laser-driven population transfer from $\ket{0}$ to the metastable $D_{5/2}$ state takes place. Then, the qubits are singled at the LIZ, where state-dependent laser-induced fluorescence is detected. Thresholding the number of detected photons allows for assigning detection events to logical ($Z$) basis states, and equivalently to eigenvalues $M_i=\pm 1$ of the Pauli operator $Z_i$ of qubit $i$:  
\begin{eqnarray}
\text{`dark'} \rightarrow\ket{0} \Leftrightarrow M_i^{(Z)}=+1 \nonumber \\
\text{`bright'} \rightarrow\ket{1} \Leftrightarrow M_i^{(Z)}=-1
\end{eqnarray}
Logical results on rotated bases $M_i^{(X)}=\pm 1$ ($M_i^{(Y)}=\pm 1$) are acquired by performing an analysis rotation $R(\pi/2,-\pi/2)$ ($R(\pi/2,0)$) on the respective qubit before shelving and fluorescence detection. As the population transfer on all qubits is carried out before fluorescence detection, cross-talk errors throughout readout are avoided. Details on qubit and shuttling operations and the sequences for register preparation and readout can be found in the supplemental material \cite{supplemental}.

\section{Measurement Results}

\subsection{Parity readout in the logical basis}
\begin{figure}[h!tp]\begin{center}
\includegraphics[width=\columnwidth]{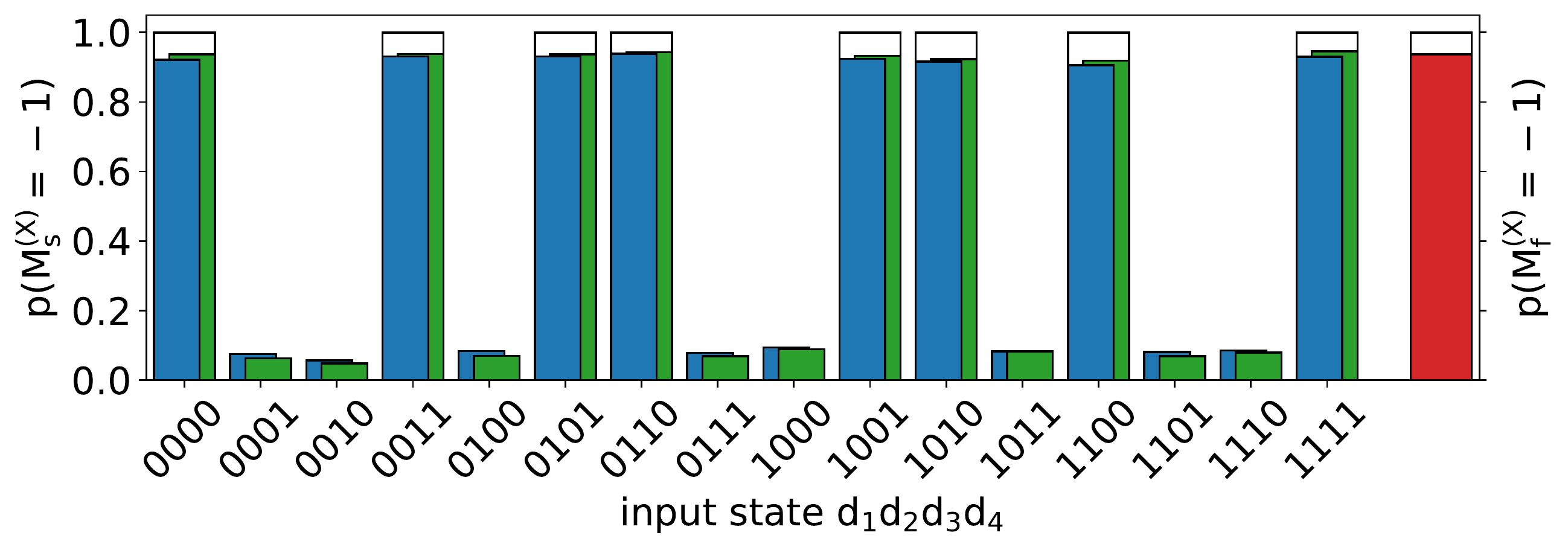}
\caption{Fault-tolerant parity readout. The syndrome $M_s^{(X)}=-1$ event rate is shown for each computational basis input state of the data qubits, for all valid shots (blue) and post-selected on the flag qubit (green), versus ideal rate (white). 960 shots per input state are measured, the average shot noise error per input state is about 7$\times 10^{-3}$. The flag $M_f^{(X)}=-1$ readout rate is shown separated at the right (red). }
\label{fig:6IonFTR}
\end{center}
\end{figure}
We first verify the functionality of the FT PCM protocol by carrying out the sequence shown in Fig. \ref{fig:6IonFTRShuttling}, while preparing the data qubits in all 16 computational basis states. Syndrome and flag qubit are initialized to $\ket{-} = \frac{1}{\sqrt{2}}(\ket{0}-\ket{1})$ by means of an $R(\pi/2,-\pi/2)$ rotation on the initial state $\ket{0}$. The measurement results $M_s^{(X)}$ of the syndrome are compared to the parity of the input state. We define the parity fidelity as
\begin{eqnarray}
    \mathcal{P}&=&\frac{1}{2}\big[p(M_s^{(X)}=-1\;|\;P_{in}=+1) \nonumber \\
    &&+p(M_s^{(X)}=+1\;|\;P_{in}=-1)\big],
\end{eqnarray}

i.e. the probability for the correct syndrome readout result $M_s^{(X)}$ conditioned on the input parity $P_{in}$ of the data qubits. For 960 shots per input state, we measure $\mathcal{P}=$~92.3(2)\%, see Fig.~\ref{fig:6IonFTR}. For 93.7(2)\% of all shots, the flag qubit is detected as $M_f^{(X)}=-1$, indicating a low rate of weight-2 errors. Post-selecting the syndrome measurement on the flag readout, we obtain a conditional parity fidelity of  $\mathcal{P}=$~93.2(2)\%. It exceeds the bare parity fidelity by 4.5 standard errors, thus showing that the FT scheme operates in the regime where it can catch native errors occurring throughout the PCM sequence. A discussion on the relevant error sources can be found in the supplemental material \cite{supplemental}.

\renewcommand{\arraystretch}{1.6}
\begin{table*}%[H]
  \centering
  \begin{tabular}{ |wc{0.7cm}|p{4cm}||wc{5.5cm}|wc{6.5cm}|  }
\hline
 &  & $n=4$ & $n=6$ \\
 \hline \hline
 1 & GME state $\ket{\psi}^{(n)}_{out}$  &  ${1\over \sqrt{2}}\left(\ket{\psi}_{out}^{(4)}\ket{-}_{s}+\ket{\psi^{\perp}}_{out}^{(4)}\ket{+}_{s}\right)$  & ${1\over \sqrt{2}}\left(\ket{\psi}_{out}^{(5)}\ket{-}_{f}+i\ket{\psi^{\perp}}_{out}^{(5)}\ket{+}_{f}\right)$ \\
 \hline
 2 & Substate $M_{s/f}^{(X)} = +1$ &  $\ket{\psi}_{out}^{(4)}={1\over \sqrt{2}}\left(\ket{--++}+\ket{++--}\right)$  & $\ket{\psi}_{out}^{(5)}={1\over \sqrt{2}}\left(\ket{--++-}+\ket{++--+}\right)$   \\
 \hline
 3 & Substate $M_{s/f}^{(X)} = -1$ &   $\ket{\psi^{\perp}}_{out}^{(4)}={1\over \sqrt{2}}\left(\ket{--++}-\ket{++--}\right)$  & $\ket{\psi^{\perp}}_{out}^{(5)}={1\over \sqrt{2}}\left(\ket{----+}+\ket{++++-}\right)$  \\
 \hline
  & & $\{  g_1 = X_1X_2, g_2=-X_2X_3,$ & $\{ g_1 = -X_3X_s, g_2=-X_4X_s, $\\
 4 & Stabilizer generator set $\mathcal{S}_n$    &  $g_3=X_3X_4, g_4=\pm Z_1Z_2Z_3Z_4\}$ &  $g_3=-X_1X_sX_f, g_4=-X_2X_sX_f, $\\
 &&& $g_5=Z_1Z_2Y_f,g_6=Z_1Z_2Z_3Z_4Z_s\}$\\
 \hline
 5 & Witness operator $W_n$    & $l_4\mathbb{1}-{1\over 4}\sum_{i=1}^{4}g_i$ & $  l_6\mathbb{1}-{1\over 6}\sum_{i=1}^{6}g_i$\\
 \hline
 6 & Bound $l_n$    & 3/4 & 5/6\\
 \hline
    \end{tabular}
    \caption{Properties of the entangled states generated by the PCM circuit for suitable input states. We distinguish the cases of the $n=4$ data qubits and all $n=6$ qubits involved in the respective GME state. The GME state vectors (before any measurements) are shown in line \textbf{1}. The corresponding substates are shown in lines \textbf{2} and \textbf{3}. The set of stabilizer generators with eigenvalues $+1$ which fix the four-qubit states $\ket{\psi}_{out}^{(4)}$ and $\ket{\psi^{\perp}}_{out}^{(4)}$, as well as the six-qubit state $\ket{\psi}^{(6)}_{out}$ are shown in line \textbf{4}, witness operators $W_n$ constructed based on these generator sets, according to Eq.~(\ref{eq:the-witness-we-use}), are displayed in line \textbf{5} together with the corresponding threshold values $l_n = (n-1)/n$ in line \textbf{6}.}
    \label{tab:GMEeqs}
\end{table*}

\subsection{Error Injection}

\begin{figure}[h!tp]\begin{center}
\includegraphics[width=\columnwidth]{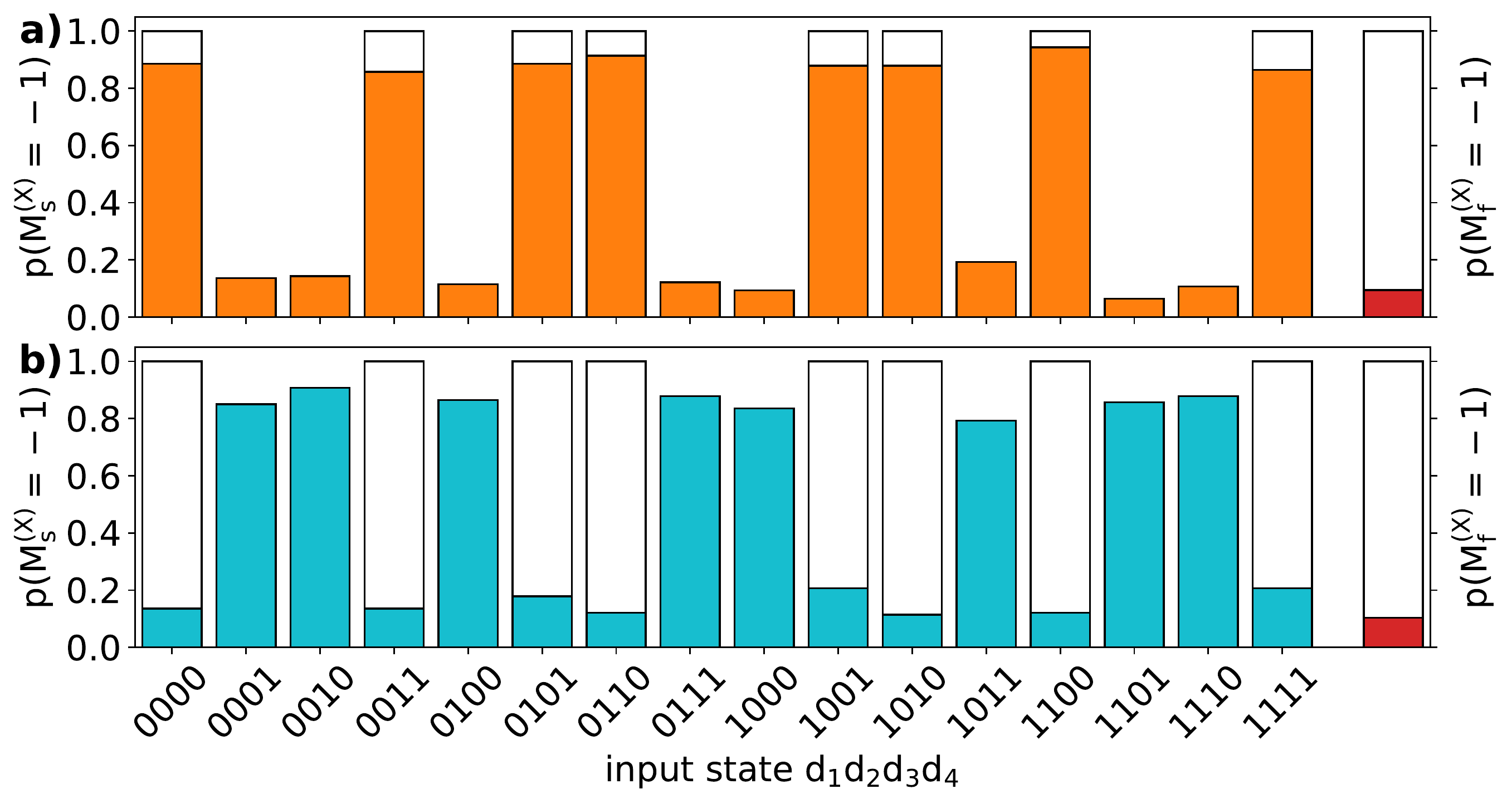}
\caption{Fault-tolerant parity readout including an injected error between gates $d_2s$ and $d_3s$. The syndrome $M_s^{(X)}=-1$ event rate is shown versus the logical input state of the data qubits. 140 shots per input state are measured, the average shot noise error per input state is about 2.5$\times 10^{-2}$. \textbf{a)} pertains to a $Y$-type error, which does not affect the syndrome readout. \textbf{b)} pertains to an $X$-type error, which also flips the logical result of the syndrome. In both cases, the flag qubit is detected predominantly in $M_f^{(X)}=+1$, corresponding to detection of the error.}
\label{fig:6IonFTR_Error}
\end{center}
\end{figure}

In order to explicitly demonstrate that the FT PCM scheme can reliably detect hook errors, we deliberately inject errors $R_s(\pi,0) \equiv X_s$ or $R_s(\pi,\pi/2) \equiv Y_s$ on the syndrome qubit (equivalence up to a global phase), between gates $d_2s$ and $d_3s$, see Fig. \ref{fig:circuit}. The resulting $M_{s,f}^{(X)}=-1$ event rates for syndrome and flag are shown in Fig. \ref{fig:6IonFTR_Error}. The injected $Y_s$, which corresponds to the simultaneous occurrence of a bit and a phase flip error, does not commute with the  subsequent entangling gates Eq. \ref{eq:entanglinggates}. The three subsequent entangling gates involving the syndrome qubit lead to a final $X_s$ error, which is not detected upon syndrome measurement. It also proliferates to the data qubits in the form of $Z_3$ and $Z_4$ errors and would therefore compromise the encoded state. However, as it propagates to the flag as a $Z_f$ error via entangling gate $sf$, it can still be detected. We observe an error detection rate of 90.6(6)\% $M_f^{(X)}=+1$ events on the flag. The syndrome still corresponds to the logical input state's parity with $\mathcal{P}=$~88.3(7)\%. By contrast, the injected $X_s$ error results in a final $Y_s$ error, such that the final syndrome $M_s^{(X)}=-1$ events anti-correlate with the input parity, yielding $\mathcal{P}=$~14.7(7)\%. Thus, as expected, the injected error results in a PCM measurement error. Similarly to the $Y_s$ error, an error detection rate of 89.7(6)\% $M_f^{(X)}=+1$ events is observed on the flag qubit. The latter indicates that the flag qubit again reliably detects the propagation of potentially detrimental weight-2 $Z$ errors onto the data qubits, as is required to preserve the fault-tolerance of the scheme.

\subsection{Generation of Genuine Multipartite Entanglement}
The PCM scheme can be used to generate maximally entangled states. We verify four-qubit GME involving only the data qubits and extend it to the case of six-qubit GME, including the syndrome and flag qubits. The verification of $n$-qubit GME is carried out efficiently via measurement of witness operators \cite{GUEHNE09,BOUR04}, as the measurement overhead for complete state tomography \cite{ROOS04} would scale detrimentally with $n$.

In contrast to the measurements discussed before, now all data qubits are initialized in $\ket{+}={1 \over \sqrt{2}}(\ket{0}+\ket{1})$ via local rotations $R_{d_i}(3\pi/2,-\pi/2)$ applied to $\ket{0}$. The GME states generated by the PCM 
%are equivalent to GHZ states up to local rotations and 
are listed in Table \ref{tab:GMEeqs} (lines 1-3). The GME states are verified via entanglement witnessing~\cite{GUEHNE09,Friis2019}. %,Toth2005}.
 An entanglement witness $W$ is an observable whose expectation value is by construction positive or equal to zero for all separable states, $\text{tr}(W\rho) \geq 0$, and negative for specific entangled states, $\text{tr}(W\rho) < 0$. The four- and six-qubit output states $\ket{\psi}^{(n)}_{out}$ of the PCM circuit belong to the class of stabilizer states, for which we use entanglement witness operators of the form 
\begin{equation}
W_n = l_n\mathbb{1}-{1\over n}\sum_{i=1}^{n}g_i
\label{eq:the-witness-we-use}
\end{equation}
with the constant $l_n = (n-1)/n$. These witnesses correspond -- up to a normalization factor $1/n$ -- to the witnesses proposed in Refs.~\cite{Toth2005,PhysRevLett.94.060501}. They can be efficiently evaluated as they require the measurement of only the $n$ stabilizer generators $g_i$ (see Table \ref{tab:GMEeqs}, line 4). The expected ideal output states can be uniquely defined as eigenstates of the stabilizer generators with eigenvalue $+1$: 
\begin{equation}
g_i \ket{\psi}^{(n)}_{out} = + \ket{\psi}^{(n)}_{out}
\end{equation}
Thus, GME in the experimentally prepared $n$-qubit states is signalled by a negative witness expectation value, $\langle W_n \rangle < 0$, which is the case if the sum of generator expectation values $\frac{1}{n}\sum_{i}^{n}g_{i}$ exceeds the threshold value of $l_n=(n-1)/n$, amounting to thresholds of $l_4=3/4$ and $l_6 =5/6$ for the verification of four- and six-qubit GME, respectively. Each generator expectation value $\langle g_i\rangle$ is determined by measuring the qubit register in a measurement setting where each qubit $j$ is subjected to appropriate analysis pulses before readout, which feature drive phases corrected for systematic phases acquired throughout the PCM sequence.

\begin{figure}[h!tp]\begin{center}
\includegraphics[width=0.9\columnwidth]{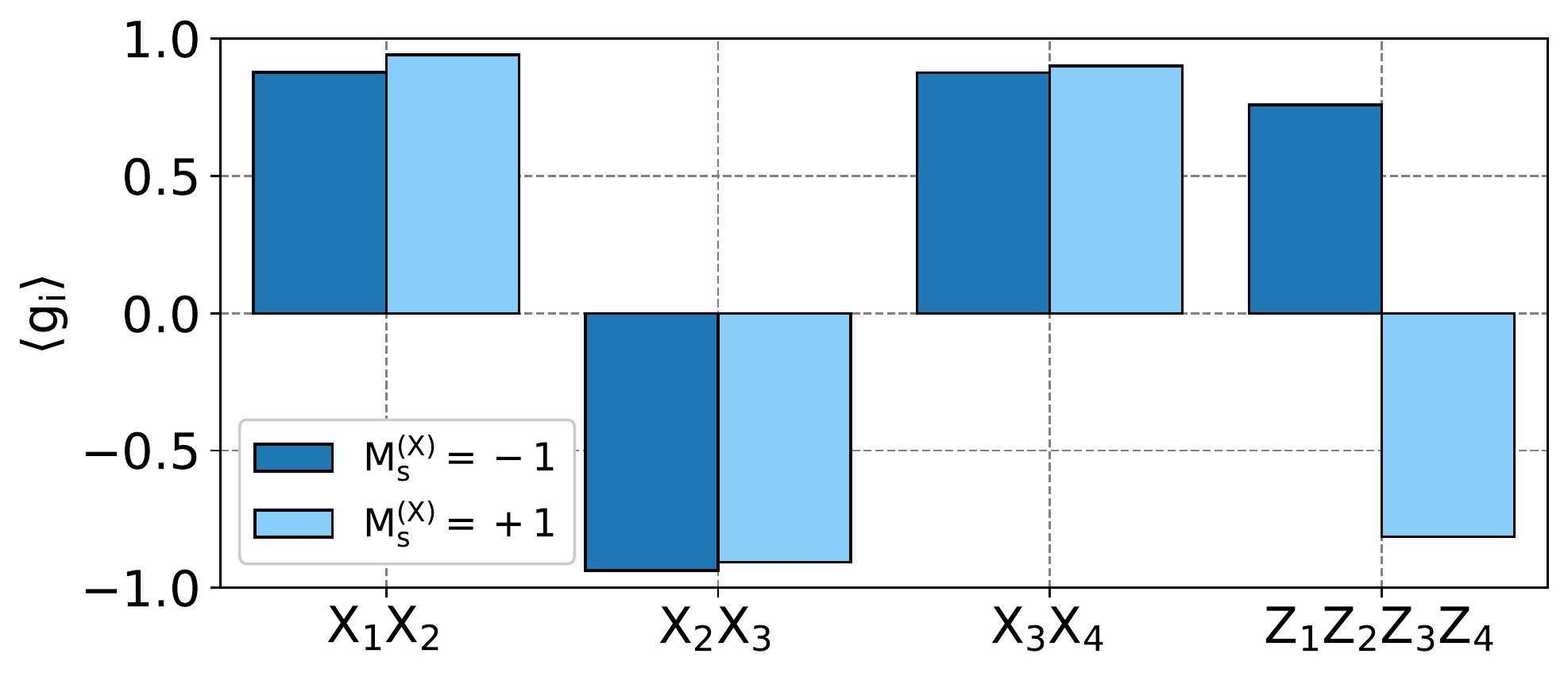}
\caption{Expectation values of stabilizer generators of the four-qubit GME state. About 330 per $X$-type stabilizer and 990 shots for the $Z$-type stabilizer are acquired. The results are conditioned on the readout result $M_s^{(X)}$ of the syndrome. The average shot-noise errors are about 2$\times 10^{-2}$ for all stabilizer expectation values. }
\label{fig:4QBGME}
\end{center}
\end{figure}

%4QB text 
For preparation of a four-qubit GME state, the entangling gates $sf$ between the syndrome and ancilla qubit are switched off, however the respective rephasing pulses are retained. The measured expectation values of the four stabilizer generators defined on the data qubits are shown in Fig.~\ref{fig:4QBGME}, each conditioned on the syndrome measurement result $M_s^{(X)}$. Consistently with the four-qubit GME state in Table \ref{tab:GMEeqs}, $X_1X_2,X_3X_4$ ($X_2X_3$) display even (odd) parity, respectively. The parity of $Z_1Z_2Z_3Z_4$ depends conditionally on the syndrome readout result $M_s^{(X)}$, as upon syndrome readout, the four data qubits are projected into the respective $\pm 1$ eigenspace of the measured parity check operator. For the input product state $\ket{+}^{\otimes 4}$ of the data qubits, they are projected into one of two four-qubit GHZ-type states. As the substates of the data qubits forming the GME state feature opposite parities in the $X$ basis, this shows that the PCM circuit reliably measures the parity in this basis upon initial basis change of the data qubits. Evaluating the entanglement witness expectation value conditionally on the syndrome measurement result, we obtain $\expec{W_{4}}=-0.14(1)$ for $M_s^{(X)}=+1$ and $\expec{W_{4}}=-0.11(1)$  for $M_s^{(X)}=-1$, respectively. Both values fall below zero by more than 10 standard errors, we therefore certify conditional four-qubit GME.

\begin{figure}[h!tp]\begin{center}
\includegraphics[width=\columnwidth]{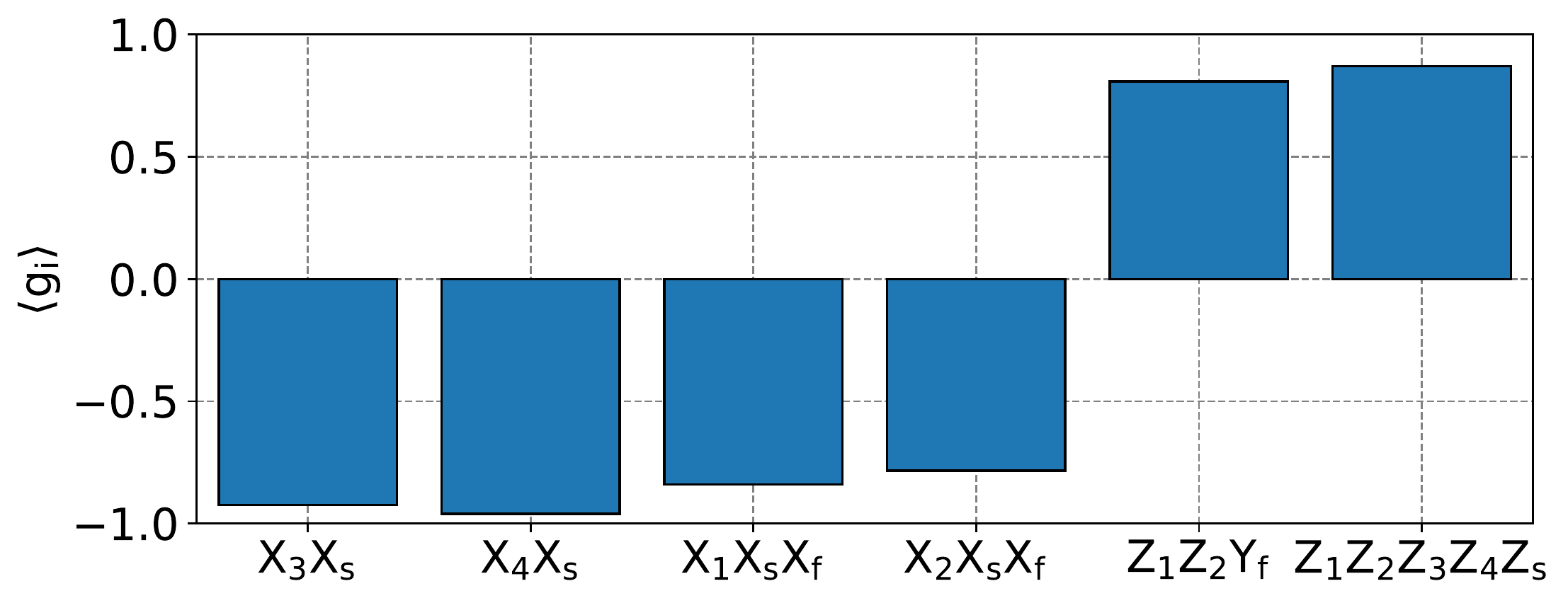}
\caption{Expectation values of stabilizer generators $g_i$ for verification of six-qubit GME. Each $X$-type stabilizer is evaluated from 500 shots, while the $Z$-type stabilizers are evaluated from 1000 shots each. The average shot-noise errors are about 2$\times 10^{-2}$ for all stabilizer expectation values.}
\label{fig:6QBGME}
\end{center}
\end{figure}

%6QB text 
For the generation of six-qubit GME, we introduce additional rotations on the syndrome qubit, $R_s(\pi/2, 0)$, between the $d_2s$ and $d_3s$ entangling gates \cite{PRXQuantum.2.020304} and $R_s(3\pi/2, -\pi/2)$ directly before the analysis rotation. Note that the first of these rotations on the syndrome qubit can be interpreted in the QEC context as a coherent rotation error, which propagates through the  subsequent gates and results in a six-qubit equal-weighted coherent superposition state. Here, the first component corresponds to state where the error has propagated into two data qubit errors, captured by the flag-qubit (in $\ket{-}_f$), and the second corresponds to the fault-free component (in $\ket{+}_f$). The measured stabilizer expectation values are shown in Fig. \ref{fig:6QBGME}, from which we compute the expectation value of the witness $W_6$ (see Table \ref{tab:GMEeqs}, line 5), obtaining $\expec{W_{6}}=-0.031(8)$. Falling below zero by 3.8 standard errors, we certify the capability of the flag-based PCM circuit to generate a six-qubit GME state.

\section{Discussion and Outlook}
\begin{figure}[h!tp]\begin{center}
\includegraphics[width=\columnwidth]{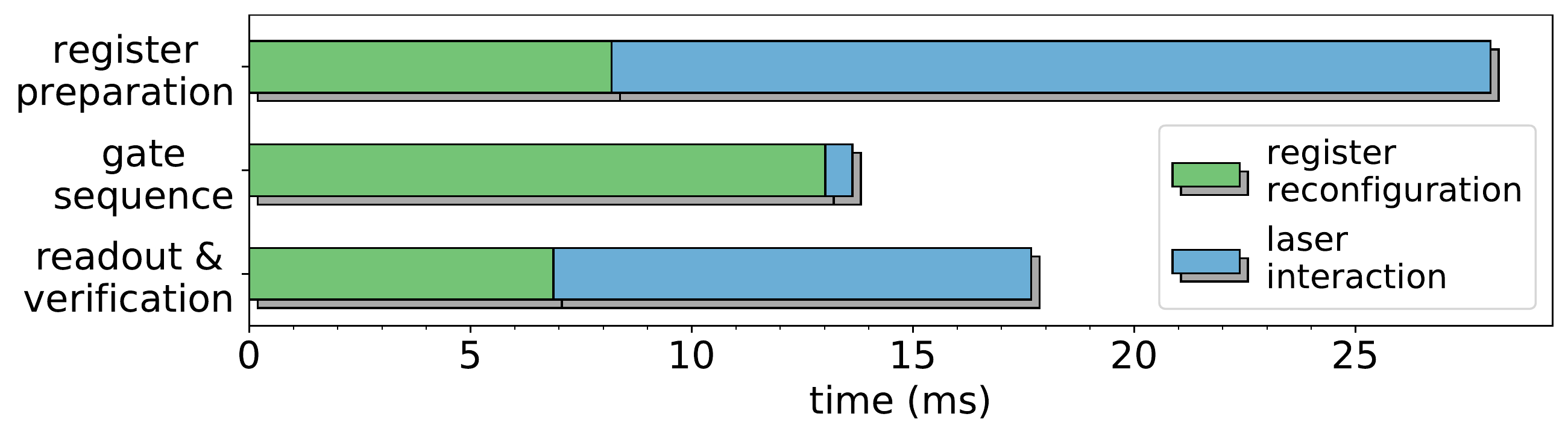}
\caption{Timing budget of the FT PCM sequence. The sequence is subdivided into register preparation, the actual PCM gate sequence shown in Fig. \ref{fig:6IonFTRShuttling}, and the readout and verification of all qubits. For each block we show the timing overhead for shuttling operations (green) and laser-driven gate operations (blue).}
\label{fig:6IonFTRTiming}
\end{center}
\end{figure}
We have successfully demonstrated a low qubit-overhead FT PCM scheme on a shuttling-based trapped-ion quantum processor. To that end, we have verified a parity measurement with high single-shot fidelity, that increases when taking the flag ancilla into account. By introducing errors deliberately, we have shown that the flag ancilla reliably detects the occurrence of errors, which would otherwise proliferate into uncorrectable weight-2 errors on the data qubit register. This verifies the FT operation of the PCM as key building block for QEC protocols.
Furthermore, we have efficiently and holistically benchmarked the proper operation of the FT PCM scheme by witnessing four- and six-qubit GME generation for suitable input states. A key enabling feature of the FT PCM scheme is the virtually complete absence of crosstalk errors during gate operations. Beyond their relevance in the context of FT QEC, our results demonstrate the capability of realizing multi-qubit quantum protocols with effective all-to-all connectivity on a shuttling-based quantum processor architecture. 

The timing budget of the protocol is shown in Fig. \ref{fig:6IonFTRTiming}. About 23\% of the duty cycle pertains to the PCM gate sequence, while the remainder pertains to register initialization -- mostly cooling -- and fluorescence readout. Of the the actual PCM gate sequence, 5\% of the execution time is given by laser-driven gate interactions, while 95\% is consumed by shuttling operation overhead. While the preparation and readout overheads scale linearly with the register size, the shuttling overhead pertaining to the gate sequences can scale up to quadratically with the qubit number, depending on the connectivity required by the underlying protocol. This overhead can be mitigated by improving the control hardware to decrease the time required for qubit register reconfigurations, but also by using multiple manipulation sites for parallel processing \cite{Mehta2020}. We have carried out a circuit of moderate depth \textit{without} the requirement of in-sequence cooling for removing excitation of vibrational modes of the ions incurred from the shuttling operations. Future extensions will include dual-species operation for sympathetic cooling for further increase of sequence depths, and for non-destructive in-sequence readout \cite{Ballance2015,Tan2015}. These improvements will pave the way for using the FT PCM demonstrated here as a building block for FT QEC cycles executed on complete logical qubits \cite{PhysRevX.7.041061,PhysRevA.100.062307}.

\begin{acknowledgments}
	We acknowledge former contributions of Thomas Ruster, Henning Kaufmann and Christian Schmiegelow to the experimental apparatus. The research is based upon work supported by the Office of the Director of National Intelligence (ODNI), Intelligence Advanced Research Projects Activity (IARPA), via the U.S. Army Research Office grant W911NF-16-1-0070. The views and conclusions contained herein are those of the authors and should not be interpreted as necessarily representing the official policies or endorsements, either expressed or implied, of the ODNI, IARPA, or the U.S. Government. The U.S. Government is authorized to reproduce and distribute reprints for Governmental purposes notwithstanding any copyright annotation thereon. Any opinions, findings, and conclusions or recommendations expressed in this material are those of the author(s) and do not necessarily reflect the view of the U.S. Army Research Office. We gratefully acknowledge funding from the EUH2020-FETFLAG-2018-03 under Grant Agreement no.820495, the European Research Council (ERC) via ERC StG QNets Grant Number 804247, and by the Germany ministry of science and education (BMBF) via the VDI within the projects VERTICONS and IQuAn.
\end{acknowledgments}	
%	\clearpage\newpage

	\bibliographystyle{apsrev4-1}
	\bibliography{lit_etal}

\clearpage
%\includepdf[pages=-,pagecommand={},width=\textwidth]{./figs/6QBFTRsupp}
\includepdf[pages={{},1,{},2,{},3,{},4,{},5}]{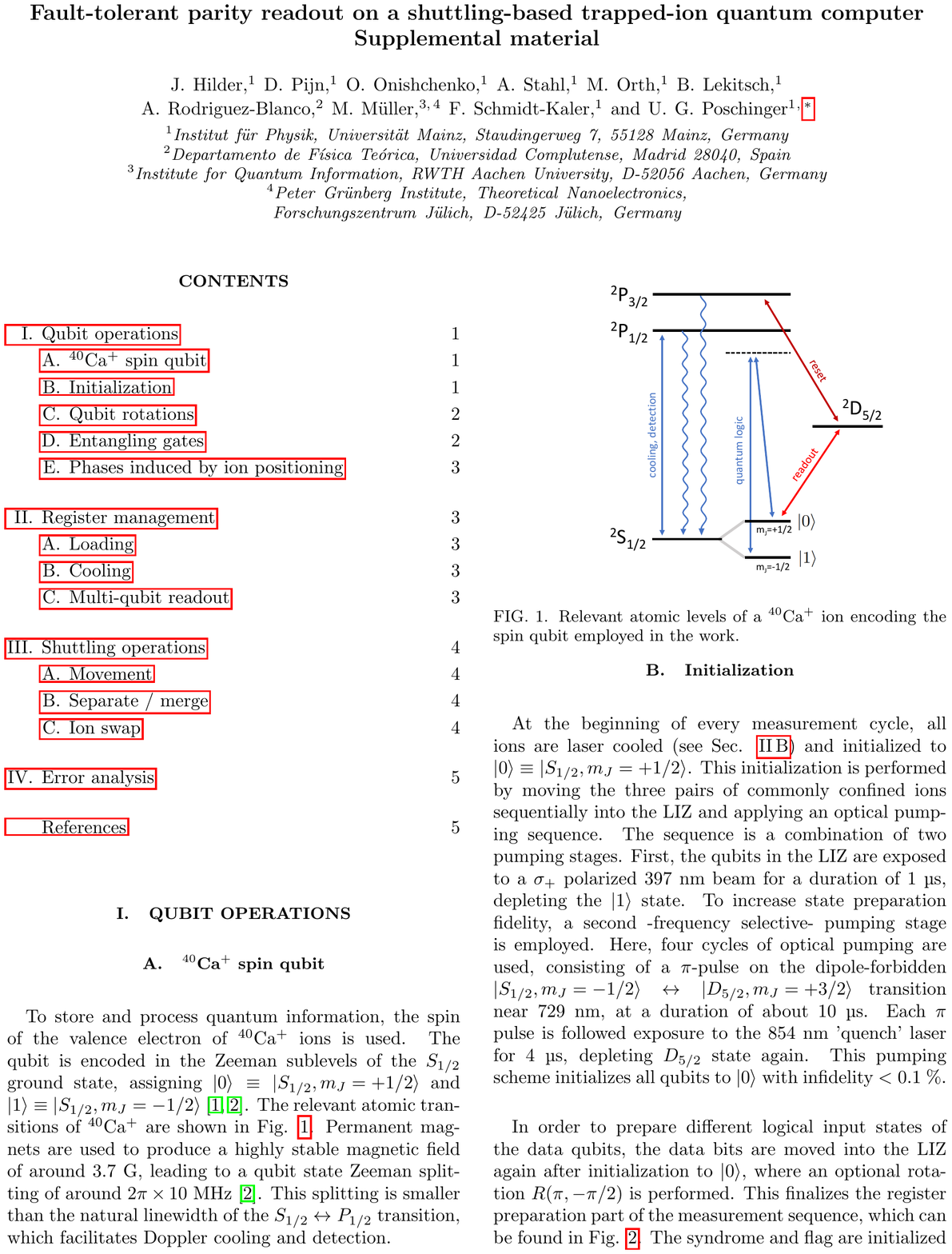}

\end{document}

\title{Fault-tolerant parity readout on a shuttling-based trapped-ion quantum computer \\ Supplemental material}

\author{J.~Hilder}
\affiliation{Institut f\"ur Physik, Universit\"at Mainz, Staudingerweg 7, 55128
	Mainz, Germany}
\author{D.~Pijn}
\affiliation{Institut f\"ur Physik, Universit\"at Mainz, Staudingerweg 7, 55128
	Mainz, Germany}
\author{O.~Onishchenko}
\affiliation{Institut f\"ur Physik, Universit\"at Mainz, Staudingerweg 7, 55128
	Mainz, Germany}
\author{A.~Stahl}
\affiliation{Institut f\"ur Physik, Universit\"at Mainz, Staudingerweg 7, 55128
	Mainz, Germany}
\author{M.~Orth}
\affiliation{Institut f\"ur Physik, Universit\"at Mainz, Staudingerweg 7, 55128
	Mainz, Germany}
\author{B.~Lekitsch}
\affiliation{Institut f\"ur Physik, Universit\"at Mainz, Staudingerweg 7, 55128
	Mainz, Germany}
\author{A.~Rodriguez-Blanco}
\affiliation{Departamento de F\'isica Te\'orica, Universidad Complutense, Madrid 28040, Spain}
\author{M.~M\"uller}
\affiliation{Institute for Quantum Information, RWTH Aachen University, D-52056 Aachen, Germany}
\affiliation{Peter Gr\"unberg Institute, Theoretical Nanoelectronics,Forschungszentrum J\"ulich, D-52425 J\"ulich, Germany}
\author{F.~Schmidt-Kaler}
\affiliation{Institut f\"ur Physik, Universit\"at Mainz, Staudingerweg 7, 55128
	Mainz, Germany}
\author{U.~G.~Poschinger}\email{poschin@uni-mainz.de}
\affiliation{Institut f\"ur Physik, Universit\"at Mainz, Staudingerweg 7, 55128
	Mainz, Germany}

\maketitle
\tableofcontents

\section{Qubit operations}

\subsection{$^{40}$Ca$^+$ spin qubit}

\begin{figure}[h!tp]\begin{center}
\includegraphics[width=0.8 \columnwidth,trim={4cm 0cm 8cm 0cm},clip]{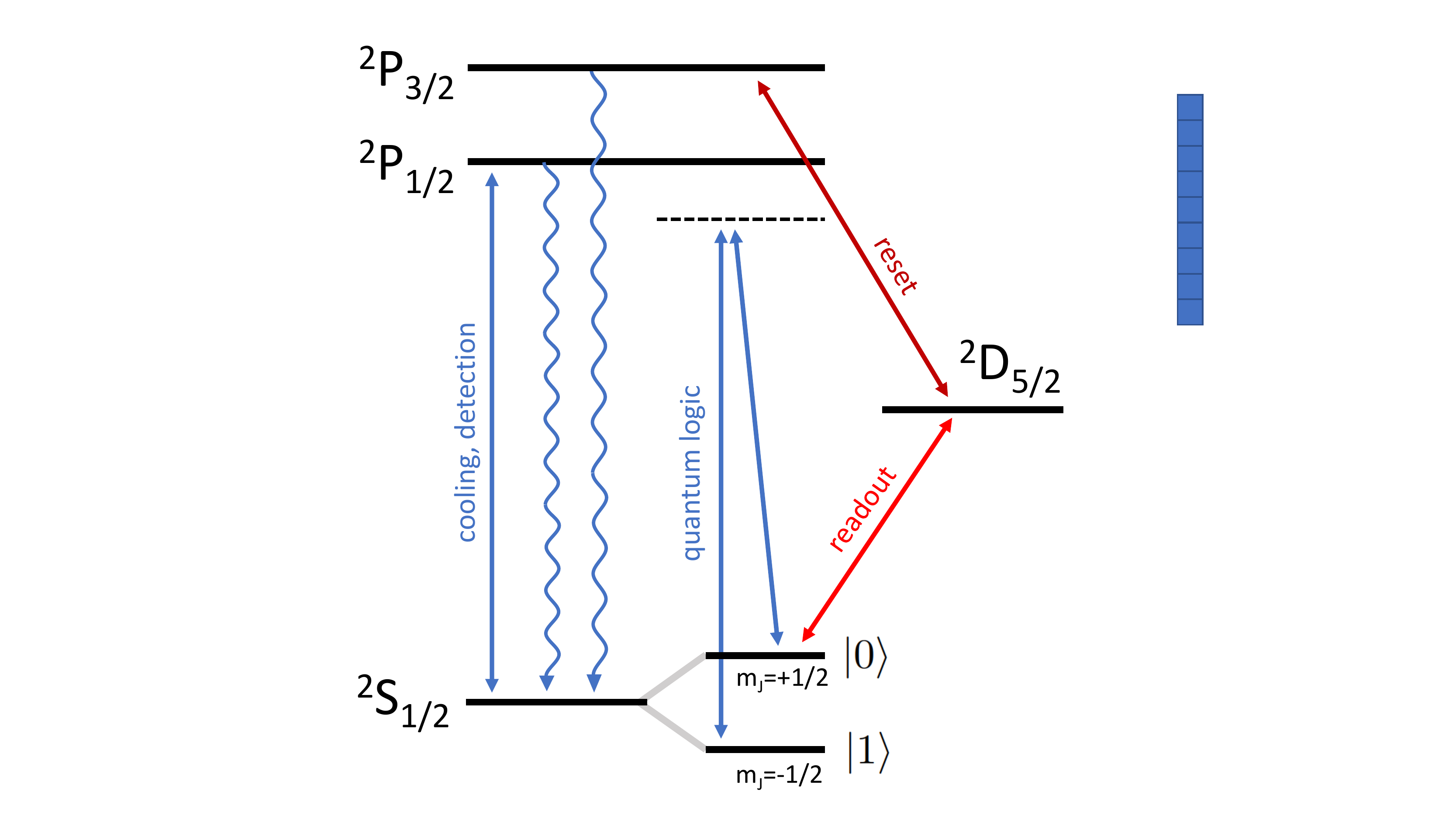}
\caption{Relevant atomic levels of a $^{40}$Ca$^+$ ion encoding the spin qubit employed in the work.}
\label{fig:levelscheme}
\end{center}
\end{figure}

\begin{figure*}[h!tp]\begin{center}
\includegraphics[width=\textwidth]{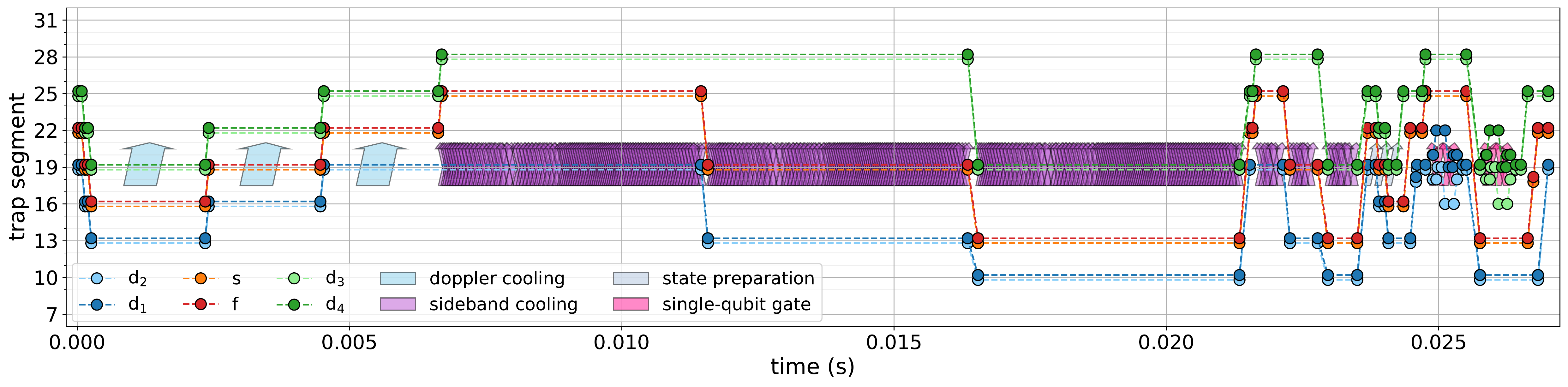}
\caption{Shuttling of the register preparation sequence part described in Sec. \ref{sec:init}. The two-qubit building blocks sequentially undergo Doppler laser cooling (light blue arrows, up to 6~ms), pulsed sideband cooling on all transverse modes (purple arrows, up to 23~ms), optical pumping (light blue arrows, at 24~ms) and optional qubit rotations for preparation of arbitrary logical states of the data qubits (pink arrows, up to 26~ms). }
\label{fig:6IonFTRShuttling_RegisterPrep}
\end{center}
\end{figure*}

To store and process quantum information, the spin of the valence electron of $^{40}$Ca$^+$ ions is used. The qubit is encoded in the Zeeman sublevels of the $S_{1/2}$ ground state, assigning $\ket{0}\equiv \ket{S_{1/2},m_J=+1/2}$ and $\ket{1}\equiv \ket{S_{1/2},m_J=-1/2}$ \cite{POSCHINGER2009,RusterLongLived2016}. The relevant atomic transitions of $^{40}$Ca$^+$ are shown in Fig. \ref{fig:levelscheme}. Permanent magnets are used to produce a highly stable magnetic field of around 3.7~G, leading to a qubit state Zeeman splitting of around $2\pi\times10$~MHz \cite{RusterLongLived2016}. This splitting is smaller than the natural linewidth of the $S_{1/2}\leftrightarrow P_{1/2}$ transition, which facilitates Doppler cooling and detection.

\subsection{Initialization}
\label{sec:init}
At the beginning of every measurement cycle, all ions are laser cooled (see Sec. \ref{sec:cooling}) and initialized to $\ket{0}\equiv \ket{S_{1/2},m_J=+1/2}$. This initialization is performed by moving the three pairs of commonly confined ions sequentially into the LIZ and applying an optical pumping sequence. The sequence is a combination of two pumping stages. First, the qubits in the LIZ are exposed to a $\sigma_+$ polarized 397~nm beam for a duration of 1~\si{\micro\second}, depleting the $\ket{1}$ state. To increase state preparation fidelity, a second -frequency selective- pumping stage is employed. Here, four cycles of optical pumping are used, consisting of a $\pi$-pulse on the dipole-forbidden $\ket{S_{1/2},m_J=-1/2}\leftrightarrow\ket{D_{5/2},m_J=+3/2}$ transition near 729~nm, at a duration of about 10~\si{\micro\second}. Each $\pi$ pulse is followed exposure to the  854~nm 'quench' laser for 4~\si{\micro\second}, depleting $D_{5/2}$ state again. This pumping scheme initializes all qubits to $\ket{0}$ with infidelity $<0.1$~\%.\\

In order to prepare different logical input states of the data qubits, the data bits are moved into the LIZ again after initialization to $\ket{0}$, where an optional rotation $R(\pi,-\pi/2)$ is performed.  This finalizes the register preparation part of the measurement sequence, which can be found in Fig. \ref{fig:6IonFTRShuttling_RegisterPrep}. The syndrome and flag are initialized to $\ket{-}$ by a single-qubit rotation $R(\pi/2,-\pi/2)$ applied at the beginning of the gate sequence. 

Readout rotation pulses $R(\pi/2,-\pi/2)$ on syndrome and flag are performed at the end of the gate sequence.

% Shelving, Detection
%Reset - quench, duration

\subsection{Qubit rotations}

Local qubit rotations are carried out by driving a stimulated Raman transition with two beams near 397~nm, with a frequency matching the qubit frequency. The drive field is detuned from the $S_{1/2} \leftrightarrow P_{1/2}$ transition by about $2\pi\times$~1~THz. The beams are co-propagating, therefore the effective wavevector is zero, and the qubit drive does not couple to the motion of the qubit ions. Single-qubit Clifford error-per-gate rates of down to 10$^{-4}$ have been measured via randomized benchmarking.

Local qubit rotations occurring after initial rotations are corrected for systematic phases. The analysis pulses on the syndrome and flag qubits are corrected by a calibrated phase, for taking into account additional systematic phases resulting from the dynamical repositioning of the qubit ions in an inhomogeneous magnetic field, see Sec. \ref{sec:sip}. For the data qubits only, the analysis pulse phases are shifted by $-\pi/2$ to take phase shifts from the entangling gates into account. For the GME generation and verification presented in main manuscript, we apply the initial $3\pi/2$ rotation on each data qubit directly before the respective entangling gate to the syndrome, and the analysis rotation directly after the gate. This way, the data qubits spend a minimum amount of time in superposition state, which mitigates errors from dephasing.

\subsection{Entangling gates}

Entanglement between two qubits is realized using a geometric phase gate. Two laser beams at around 397~nm with a red detuning of about $2\pi\times$1.0~THz from the $S_{1/2}\leftrightarrow P_{1/2}$ transition are aligned such that the effective $\Vec{k}$-vector is oriented perpendicularly to the trap axis. The beams are arranged in lin-$\perp$-lin polarization geometry, the beat pattern therefore has a polarization gradient and leads to a spin-dependent optical dipole force on the two ions. The frequency difference of the beams is tuned close to the transverse in-phase (gate) mode at $2\pi\times4.64$~MHz, up to a detuning of $\delta \approx 2\pi\times20$~kHz. This almost-resonant drive force leads to transient oscillatory excitation of the gate mode, returning to rest at a duration of $T=2\pi/\delta \approx 50~\si{\micro\second}$. A geometric phase $\Phi$ proportional to the enclosed phase space area will be acquired, which can be tuned by the laser power. To realize a maximally entangling gate, a phase of $\Phi = \pi/2$ is required. The unitary describing phase accumulation on the even parity qubit states is

\begin{eqnarray}
    ZZ_{ij}(\Phi)&=&e^{\frac{i}{2}\Phi Z_i \otimes Z_j }.
    \label{eq:entanglinggatesPhi}
\end{eqnarray}

The actual entangling gate operation consists of two gate pulses, each leading to a phase accumulation of $\Phi=\pi/4$, interspersed by an additional rephasing pulse $R(\pi, -\pi/2)$ with a typical duration of around $4~\si{\micro\second}$ after half the phase accumulation, which leads to a total gate unitary of
\begin{eqnarray}
    G_{ij}&=&ZZ_{ij}(\pi/4)R(\pi,-\pi/2)ZZ_{ij}(\pi/4).
    \label{eq:entanglinggates}
\end{eqnarray}

The gate pulses feature a Tukey-type shape, ensuring adiabatic switching of the gate interaction. The reduced bandwidth of the gate pulses leads to suppression of errors from off-resonant excitation of spectator motional modes \cite{BALLANCE2014}. Typical two-qubit gate fidelities of around 99.6(2)~\% are reached at a total gate duration of $120~\si{\micro\second}$, verified via subspace cycle benchmarking \cite{PhysRevResearch.2.013317}.

\subsection{Phases induced by ion positioning}
\label{sec:sip}
Throughout the dynamical reconfiguration of the qubits, ions are moved along the trap axis into different storage positions and acquire additional phases due to a small inhomogeneity of the magnetic field. The maximum  difference of the qubit frequency is about $2\pi\times$~7~kHz across the entire trap. Accumulated phases can be described by local $Z$ rotations, which commute with the entangling gates and therefore do not perturb these. However, the analysis rotations ought to be corrected accordingly. We calibrate the phases from additional measurements, where the respective qubit is initialized in $\ket{-}$, the shuttling sequence is carried out without executing entangling gates but with the rephasing pulses retained. Instead of the final analysis pulses, $X$ and $Y$ measurements are carried out. From the respective expectation values, we obtain the positioning-induced phases via maximum likelihood estimation. With 40 shots per operator and qubit, we obtain a phase estimation accuracy of about 0.15~rad. For the measurement with injected errors, the phase $\phi_{err}$ of the error pulses on the syndrome are calibrated by scanning the phase of a local rotation $R(\pi, \phi_{err})$ at the error position, to perform a full spin-flip on the final syndrome readout.

\section{Register management}
\subsection{Loading}
The six qubits are loaded in pairs of two commonly confined ion qubits. To that end, three potential wells are formed using the dc electrodes of the segmented trap. The wells are spaced by at least three empty segments to reduce the probability of unintentional additional loading events. $^{40}$Ca$^+$ ions are obtained by resonant two-photon ionization from an effusive beam $^{40}$Ca atoms, using two laser beams near 374~nm and 423~nm. The ions are usually trapped by applying a trapping voltage of -2.4~V at the LIZ, while potential wells in storage regions use a trapping voltage of -6~V. Upon successful trapping, the ions are moved from the loading region to a storage position, and the next empty potential well is moved into the LIZ for loading. The potential wells are cycled through the LIZ until all qubits are stored at their desired location.

\subsection{Cooling}
\label{sec:cooling}
In each measurement cycle, the ions are cooled close to motional ground state via multiple cooling stages. First, Doppler cooling is performed using the $S_{1/2}\leftrightarrow P_{1/2}$ transition near 397~nm and an exposure time of 2~\si{\milli\second} per two commonly confined ions. The ion  pairs are cooled in sequence $\{d_3,d_4\}$, $\{s,f\}$ and $\{d_2,d_1\}$. The Doppler-cooled ions are  further cooled by using pulsed resolved sideband cooling via driving the stimulated Raman transition on the red sidebands of the corresponding transverse motional modes. Each cooling pulse realizes an approximate $\pi$~pulse on the transition 
\begin{equation}
\ket{0}\ket{n} \rightarrow \ket{1}\ket{n-1},
\end{equation}
such that the phonon number $n$ of the driven secular mode is reduced. After each pulse, optical pumping using a circularly polarized laser near 397~nm at a pulse duration of 1~\si{\micro\second} resets the state as 
\begin{equation}
\ket{1}\ket{n-1} \rightarrow \ket{0}\ket{n-1},
\end{equation}
The ion pairs are sideband cooled sequentially, in order $\{d_2,d_1\}$, $\{s,f\}$ and $\{d_3,d_4\}$. First, all pairs undergo a cooling sequence of a total duration of 4~ms, covering the 2nd and the 1st red sideband of all transverse modes for both axes perpendicular to the trap axis, in-phase $2\pi\times\{3.88,4.64\}$~MHz and out-of-phase $2\pi\times\{3.57,4.37\}$~MHz. All ions which have already been cooled accumulate a small amount of excitation due to anomalous heating, mostly on the in-phase modes. This is mitigated by a second, much shorter, round of sideband cooling only on the in-phase modes, performed in the same cooling order.

\subsection{Multi-qubit readout}
\label{sec:mqreadout}
\begin{figure*}[h!tp]\begin{center}
\includegraphics[width=\textwidth]{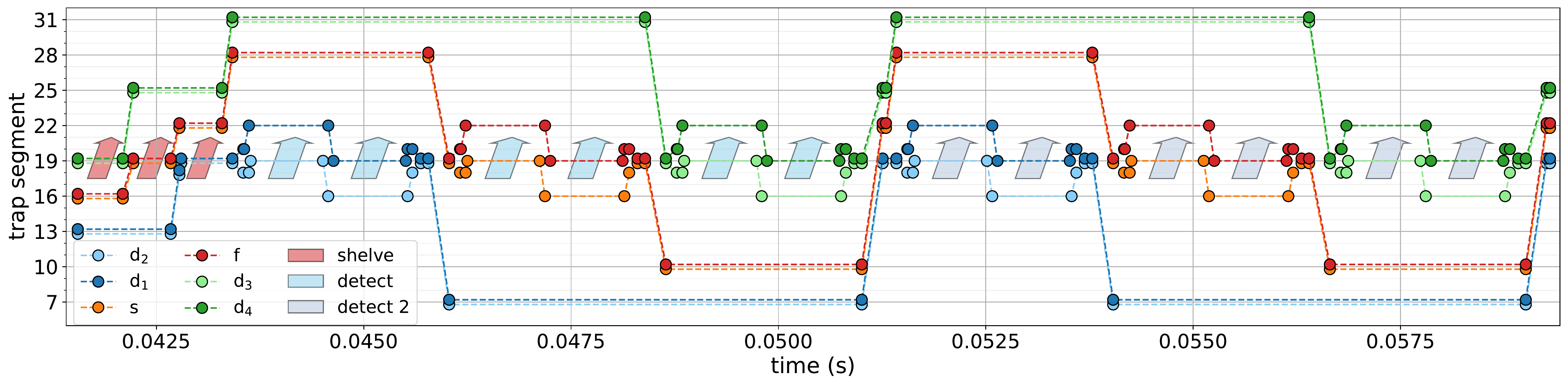}
\caption{Shuttling of the multi-qubit readout sequence part described in Sec. \ref{sec:mqreadout}. The two-qubit building blocks sequentially undergo electron shelving (red arrows, up to 43~ms). Then, state-dependent fluorescence is detected after separating each pair (light blue arrows, up to 51~ms). Another round of detection is performed including qubit reset to verify register integrity (light blue arrows, up to 59~ms). }
\label{fig:6IonFTRShuttling_Readout}
\end{center}
\end{figure*}
The sequence covers two rounds of detection for each qubit, for reading out the state of the qubit and to ensure the validity of the measurement run.  First, readout of the spin qubit requires electron shelving, i.e. selective population transfer from $\ket{0}$ to the metastable state $D_{5/2}$, which possesses a radiative lifetime of about 1~s.  This transfer is achieved by using rapid adiabatic passage pulses on the sub-transitions $\ket{0\equiv S_{1/2},m_J=+1/2}\leftrightarrow\ket{D_{5/2},m_J=+1/2}$ and $\ket{0\equiv S_{1/2},m_J=+1/2}\leftrightarrow\ket{D_{5/2},m_J=-3/2}$. For both transitions, the ions are exposed to a chirped Gaussian-shaped laser, varying the optical frequency by $\pm$60~\si{\kHz} around resonance within a duration of 200~\si{\micro\second}. The pulse parameters are chosen to maximize the transfer probability from $\ket{0}$ and to minimize parasitic transfer from $\ket{1}$. The qubits undergo electron shelving pairwise, in order $\{d_3,d_4\}$, $\{s,f\}$ and $\{d_2,d_1\}$. After pairwise shelving of all ions, a nearly identical detection sequence is executed twice, in reverse order. Here, all ion pairs are sequentially moved to the LIZ. Then, the pairs are separated, and the singled ions consecutively moved to the LIZ. They are exposed to a laser beam resonantly driving the cycling transition near 397~nm at about one-fold saturation. Scattered photons are collected and detected on a photomultiplier tube. Comparing the number of photons detected within 800~\si{\micro\second} to a threshold allows for discrimination of $\ket{0}$ and $\ket{1}$. The same detection sequence is repeated, including an additional laser beam near 854~nm, which depletes the metastable state via the  $D_{5/2}  \leftrightarrow P_{3/2}$ electric dipole transition. A complete set of 'bright' event at the second detection verifies that no ion losses have occurred during the measurement cycle. For the FT PCM shuttling sequence, we obtain a valid measurement cycle ratio of around 83\%. Finally, the ions are moved back to the initial loading positions.

\section{Shuttling operations}
\subsection{Movement}
The microstructured, segmented radio frequency ion trap consists of 32 pairs of dc electrodes,  referred to as trap segments. At most two ions are stored in one potential, to harness the lower number of motional modes in a smaller ion crystal.  In order to move ions to a neighboring segment, the voltage at the target segment is gradually put to negative bias, while the negative trapping voltage at the original segment is slowly increased at the same time. This way, the confining electrostatic potential well shifts from its original position to the destination segment. The voltage ramps are optimized to minimize the final motional excitation of the ions after the movement over a distance of one segment. The movement between two neighboring segments is performed within 20.9~\si{\micro\second}. Transport over larger segment ranges is realized by concatenated application of single-segment movements. A waiting time of 50.6~\si{\micro\second} is inserted after the last shuttling operation, before any laser-driven qubit operation is to be carried out. This ensures that the ions settle to the rest position in the LIZ.

\subsection{Separate / merge}
In order to obtain the required effective all-to-all connectivity within the six-qubit register, separation and merging of two qubit ions from / to common confinement is required. Separation is realized by dynamic control of the confining potential, transferring from a single-well to a double-well potential. In order to avoid excessive motional excitation from this operation, precise calibration of the process parameters is required. A 'tilt' voltage difference between the neighboring segments is required to compensate stray field along the trap axis, which needs to be calibrated to 1~mV precision to ensure low residual motion after the separation. This voltage is automatically recalibrated throughout data acquisition, as its value may drift due to UV light induced charge accumulation at the trap surfaces. A total separation duration of about  100~\si{\micro\second} is used. The merging process is merely the time reverse operation, employing the time-reverse voltage ramps with the same calibration parameters. The harmonic confinement along the trap axis is reduced throughout separation / merge processes, down to a minimum value of about $2\pi\times$~220~kHz.

\subsection{Ion swap}
The swap operation is realized via physical rotation of two commonly confined ion qubits, i.e. flipping the positions of the two ions along the trap axis. The rotation is controlled using the neighboring segments of the LIZ. We apply optimized voltage waveforms, minimizing the secular frequency deviation during the process to less than 300~kHz on all in-phase modes $2\pi\times\{1.49,3.88,4.64\}$~MHz and out-of-phase modes $2\pi\times\{2.57,3.57,4.37\}$~MHz of the two commonly confined ions. This avoids spectral crossing of the secular modes, and therefore suppresses transfer of motional excitation between highly excited axial modes and transverse modes. This is necessary as all transverse modes are required to have low motional excitations $\lesssim$~1 phonon in order to realize high-fidelity entangling gates. The rotation process is carried out within a duration of 60~\si{\micro\second}.

\section{Error analysis}

In this section, we discuss the relevant error sources limiting the single-shot fidelity of the parity check measurement. State preparation and measurement (SPAM) errors are of particular relevance in the context of QEC protocols: Many ancilla preparation and readout operations have to be performed per QEC cycle, and eventual feedback operations are conditioned on measurement results of these.\\
While the state preparation via two-stage optical pumping features infidelities $<$0.1\% and are thus dwarfed by measurement errors. Here, the fidelity bottleneck consists of the electron shelving operation for readout of the spin qubit. The laser beam near 729~nm used for population transfer couples to the transverse secular motion, with Lamb-Dicke factors on the four transverse modes in the ranging around $0.1$. The shuttling operations also lead to build-up of transverse motion, which impairs the electron shelving via dispersion of the coupling strength. Measurement fidelities on all qubits are listed in Table \ref{tab:SPAM}. These values are measured by initializing all qubits in logical states $\ket{0}$ ('dark' readout) or $\ket{1}$ ('bright' readout) and performing the shuttling sequence without any further gates. Readout infidelities ranging between 0.3(1)\% and 2.9(5)\% are observed, depending on qubit and prepared state. For reference, a single ion qubit without any shuttling operations before detection yields combined preparation and readout infidelities of 9(4)$\times$~10$^{-4}$ and 5(3)$\times$~10$^{-4}$ for $\ket{0}$ and $\ket{1}$, respectively. 

\renewcommand{\arraystretch}{1.4}
\begin{table}%[H]
  \centering
  \begin{tabular}{|wc{3.0cm}||wc{2.0cm}|wc{2.0cm}|}
    \hline
     FT PCM Sequence & $'bright' \equiv \ket{1}$ &  $'dark' \equiv \ket{0}$ \\
    \hline \hline
    $s$ & 98.8(2)\% & 98.6(3)\% \\
    \hline
    $f$ & 98.8(2)\% &  98.1(3)\% \\
    \hline \hline 
     GME Sequence & $'bright' \equiv \ket{1}$ &  $'dark' \equiv \ket{0}$ \\
    \hline \hline
    $d_1$ & 99.7(1)\% & 98.2(4)\% \\
    \hline
    $d_2$ & 99.6(1)\% &  98.7(3)\% \\
    \hline
    $d_3$ & 99.2(2)\% &  99.0(3)\% \\
    \hline
    $d_4$ & 98.6(2)\% &  98.6(3)\% \\
    \hline
    $s$ & 99.6(1)\% &  97.1(5)\% \\
    \hline
    $f$ & 99.6(1)\% &  98.4(3)\% \\
    \hline
    \end{tabular}
    \caption{State preparation and detection fidelity including shuttling sequence. For the data pertaining to the GME, early readout and shelving of the data qubits is carried out.}
    \label{tab:SPAM}
\vspace{0.5cm} % added spacing to increase dist to following text
\end{table}

\begin{figure}[h!tp]\begin{center}
\includegraphics[width=\columnwidth]{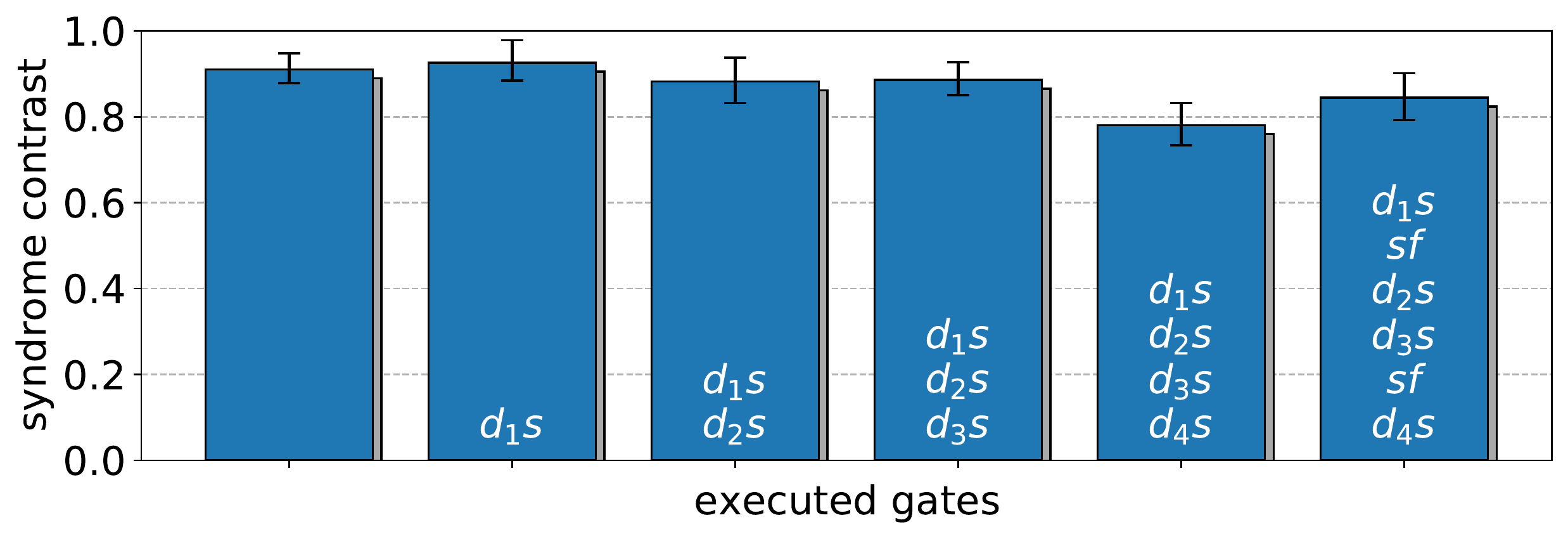}
\caption{Readout syndrome contrast after last gate with increasing number of gates executed, based on 160 shots per Pauli operator $X,Y$.}
\label{fig:ContrastLossGates}
\end{center}
\end{figure}

The infidelity of the two-qubit gates exhibits linear scaling behavior with respect to the mean phonon numbers of the transverse modes. Transverse secular motion leads to dispersion of the coupling strength of the gate driving force, which in turn leads to dispersion of the accumulated geometric phase, which finally manifests in the form of dephasing. To characterize this effect, we measure the syndrome contrast for different gates toggled on or off, performing the complete shuttling sequence. The contrast is obtained from measuring expectation values $\langle X_s\rangle$ and $\langle Y_s\rangle$ and maximum-likelihood estimation. Contrast values of ranging between 93(5)\% and 78(5)\% are observed, see Fig. \ref{fig:ContrastLossGates}. A statistically significant dependence on the number of performed two-qubit gates is not visible. We therefore conclude that this errors source does not represent a relevant contribution to the infidelity of the parity check measurement. The minimum observed contrast loss is due to fluctuations of the ambient magnetic field, and consistent with the observed parity measurement fidelity of 93.2(2)\%.

\bibliographystyle{apsrev4-1}
	\bibliography{lit_etal}